\documentstyle{article}

\begin{document}
\begin{flushright}

ITP-NENU\hspace{0.2cm}94-15

\end{flushright}

\begin{center}
\large

{\bf Quantum Dynamical Approach of Wavefunction Collapse
in  Measurement Process  and Its Application to Quantum
Zeno Effect $^*$}

\vspace{0.5cm}
  Chang-Pu Sun\\
{\it Institute of Theoretical Physics, Northeast Normal University,
Changchun 130024,\hspace{1cm} P.R.China}

\vspace{1cm}
Abstract                            \\
\end{center}
\normalsize

The systematical studies on the dynamical approach of wavefunction
collapse in quantum measurement are reported in this paper
based on the Hepp-Coleman's  model  and its generalizations. Under  certain
physically reasonable conditions, which are easily satisfied by the practical
problems,
it is shown that the off-diagonal elements of the reduced density
matrix vanish in quantum mechanical evolution process
in the macroscopic limit with a very large particle number
N. Various examples with detector made up of oscillators of
different spectrum distribution
are used to illustrate this  observations .
With the two-level system as an explicit illustration, the  quantum
information entropy is
exactly obtained to  quantitatively describe the degree of decoherence
for the  so-called partial coherence caused by detector.
The entropy for the case with many levels  is computed based on perturbation
method in the limits with very large and very small N.
As an application of this general approach for quantum measurement,
a dynamical realization of the quantum Zeno effect are present to
analyse its recent testing experiment in connection with a description
of transition in quantum information entropy. Finally, the Cini's model
for the correlation between the states of the measured system and  the detector
is generalized for the case with many energy-level.
It is shown that this generalization  can also be invoked to
give the dynamical realization of wavefunction collapse.
 \\\\\\
PACS number(s): 0365-w, 0380+r.\\
$^*$Contributed to ``Drexel International Workshop on Quantum-Classical
Correspondence", Phidelphia,Octeber, 1994.

\newpage

{\bf 1. Introduction }\\

Though quantum mechanics has been experimentally proved to be
 a quite successful
theory, its interpretation is still an important problem
that  physicists should  face [1-4]. To interpret the physical meaning of
its mathematical formalism, one has to invoke  the wave packet collapse(WPC)
(or wavefunction collapse)postulate  as an extra assumption added
to the  closed system of laws in quantum mechanics . This postulate is
also called von Neumann's projection rule or wavefunction reduction process.
Let us now describe it briefly. It is well known in quantum physics that, if
measured quantum system S is in a state $|\phi>$ that is
a linear superposition of the
eigenstates $|k>$  of the  operator $\hat A$ of an observable A just before a
 measurement, ie.,
$$|\phi>=\sum c_k|k>,c_k's ~~are~~ complex~~ numbers\eqno{(1.1)}$$
then a result of th measurement of A is one $a_k$ of
the eigenvalues of $\hat A$ corresponding to $|k>$
with the probability $|c_k|^2$. The von Neumann's postulate tell us that, once
a well-determined result $a_k$ about A has been obtained, the state of S is
no longer  $|\phi>$ and it must collapses into $|n>$ since the
immediately-successive measurement of A after the first one
should repeats the same result. In terms of the density matrix
$$\rho =|\phi><\phi|=\sum_{k,k'} c_kc_{k'}^*|k><k'|,\eqno{(1.2)}$$
for the state $|\phi>$, the above WPC process can be  expressed as
a projection or  reduction
$$\rho\rightarrow \hat {\rho}=\sum |c_n|^2|k><k|.\eqno{(1.3)}$$
Because the off-diagonal elements represent coherence, through
which the density matrix describes a non-classical probability,
the wavefunction collapse characterized by vanishing of off-diagonal
elements means the loss of quantum coherence or called quantum
decoherence [5].

The recent studies on  the quantum decoherence
in an open system $S$ surrounded by an environment $E$ was mainly motivated
by the interests in the macroscopically-quantum effects
such as dissipation in the
quantum tunnelling and the semiclassical gravity theory for particle creation
in
cosmology [6-10].
In quantum dissipation theory,
an important treatment for quantum decoherence is  invoking  quantum Brownian
motion through its  master equation to describe the vanishing of the
off-diagonal elements of reduced density matrix of $S$ [1].
The recent investigations in this context were carried out with the
Feynman-Vernons integral method for the Ohmic, sub-Ohmic or super-Ohmic
environment (e.g, see ref.[8]). Without the use of path-integral [9,10],
an exactly-solvable dynamical
model of quantum dissipation was presented by Yu
and this author to deal with the similar phenomenon.
Zurek and his collaborators especially emphasized the role of environment
surrounding the open system, which monitors the observables of the system
so that their eigenstates continuously decohere and then approach  classical
states [4,8].

It should be noticed that there exists another dynamical theory
based on Hepp-Colemen's  investigation to dynamically realize
quantum decoherence characterized by wavefunction collapse [11,12].
This theory and its generalization [13-15] were proceeded in
a purely-quantum mechanical framework.
In fact, to realize the WPC, the external classical measuring
apparatus detector must be used to detect the result. Then, someone
thinks the WPC postulate to be not quite satisfactory since quantum mechanics
is expected to be an universal theory valid for whole `universe' because the
detector, as a part of the universe, behaves classically in
the von Neumann's postulate.
A reasonable description of the detector should
be quantum essentially and it exhibits the classical or macroscopic
features in certain
limits. If one deal with the detector as a subsystem of the closed system
(the universe C = the measured system S + the detector D), it is possible
that the quantum
dynamics of the universe can result in the WPC through the interactions
between  S and D. Up to new, some exactly-solvable models have been
presented to analyse this problem. Among them, the Happ-Coleman (HC)
model is very famous
one and has been extensively studied in last twenty years [13-19].
In order to describe  studies in this paper clearly, we need to see
some details of this model.

In the original HC model, an  ultrarelativistic particle is
referred to the measured system S while a one-dimensional array of scatterers
with spin-1/2 to the detector D. The  interaction between S and D
is represented
by an homogeneous coupling

$$ H_I=\sum_{n=1}^NV( x-a_n)\sigma_1^{(n)}\eqno{(1.4)}$$
where $\sigma_1^{(n)}$ is the first component of Puli matrix; $a_n$ is the
position of the scatterer assigned to the n'th site in the array.
The Hamiltonian for D is
$$H_s=c\hat P\eqno{(1.5)}$$
 where $c$, $\hat P$ and $ x$ are the light speed, the momentum and
 coordinate operators respectively for S.
This model is quite simple, but it can be exactly solved to produce a
deep insight on the dynamical description of the quantum measurement process.
Starting with the initial state
$$|\psi(0)>=\sum c_k|k>\otimes |D>\eqno{(1.6)}$$
where $|D>$ is  pure state of D (it is usually taken to be ground state),
the evolution state $|\psi(t)>$ for the universe C=S+D is defined by the
exact solution to this model. Then, the reduced density matrix

$$\rho_s(t)=Tr_D(|\psi(t)><\psi(t)|)\eqno{(1.7)}$$
of the measured system is obtained by taking the trace of the
density matrix
$$\rho(t)=|\psi(t)><\psi(t)|\eqno{(1.8)}$$
 of the universe to the
variables of D. Obviously, $\rho_s(t)$ depends on the particle number N of
D. When $N\rightarrow \infty$, i.e., in the macroscopic limit ,
$\rho_s(t)\rightarrow \hat{\rho}$
 after long enough time t as  eq.(1-2). Namely,
the Schrodinger evolution of the
universe C=S+D leads to the WPC for the measured system.
More recently, the original CH model was improved to describe the energy
exchange between S and D by adding a free energy Hamiltonian
$$H_0=\hbar\omega \sum_{n=1}^N\sigma_3^{(n)}\eqno{(1.9)}$$
and correspondingly improving the interaction slightly. Notice that
the improved model remains exactly-solvable [13].

Because the spin quantum number is fixed to be 1/2 in the original
CH model or its improved versions, they can not
describe the {\it classical characters } of the measurement. Usually,
the classical
feature of a quantum object is determined by taking certain value for
some internal quantum numbers of the detector D or $\hbar=0$. In the case of
the
angular momentum, this classical limit corresponds to infinite spin.
 This generalized dynamical model was successfully built
by this author in 1993 [14, 15]. The first step is to establish such
a generalization of the HC model manifesting the WPC as the dynamical
process in the  classical limit as well as in the the macroscopic limit
simultaneously. Then, the essence  for
this model substantially
resulting in the realization of the WPC as a quantum dynamical process
as well as for those well-established  was found to be the factorization
of the evolution matrix in the interaction picture
with help of a  detailed study on the dynamics of
the  generalized HC model
in both the exactly-solvable case and the non-solvable case,
 For the latter,
 the high-order adiabatic approximation (HOAA) method [20-23] is applied
 to its special case that the coupling parameter depends on
the position of the measured ultrarelativistic particle quite slightly.
Finally, we point out that this possible essence in the dynamical realization
of the WPC,
is largely independent of the concrete forms of model Hamiltonians.
Notice that, in the dynamical models of wavefunction collapse for
quantum measurement, both  the macroscopic and  classical
measuring apparatus can be regarded as the environment in the certain model
of the quantum decoherence for quantum dissipation [4,8]. This is because
both they act as  the classical or macroscopic monitors seeing the system.

However, because all of the previous dynamical models of decoherence for
quantum measurement depend on the specific forms of Hamiltonians of
$D$ and $S$, it is necessary   to  present  a model-independent dynamical
approach for decoherence in quantum measurement process based on
the  HC model. It is expected that such an  approach
does not depend on the detailed structures of the Hamiltonians of $S$ and $D$
as fully as possible,  but can be invoked to deal with the practical
problems of quantum measurement such as quantum Zeno effect (QZE). This
universal  approach
should also be used to describe the role of environment in decoherence
for quantum dissipation.  It is shown in present study that, through
a suitable choice of the
interaction between $S$ and $D$, the Schrodinger evolution of the universe
$C$ formed by $S$ plus
$S$ may result in the phenomenon of decoherence in the reduced density
matrix $\hat{\rho}$
of $S$ at the macroscopic limit of $D$ with very large $N$. Mathematically,
the mechanism of this phenomenon is  that the accompanying factors in
the off-diagonal elements of reduced density matrix
$\hat{\rho}_s$ caused by the dynamical evolution of $C$ will vanish as N
approaches  infinity. It even was described in concrete examples [14,15].
Notice again that in the previous models for the quantum
measurement and decoherence, the considered systems $S$ usually are specified
as an ultrarelativistic particle or  a two-level system while the detector
$D$ as a spin array.
Here, what we require is only that the system is of the
non-degenerate discrete
spectrum and the interaction between $D$ and $S$ is  chosen to
result in a factorizable evolution matrix. However, $D$ is required
to satisfy a condition that  any row or column of
each factor corresponding to
the factorizable evolution matrix at least has one non-vanishing
off-diagonal  element. This means that the back action of the measured
system can effectively act on the detector so that the microscopic
states can be read out from the macroscopic counting numbers contributed
by all particles in detector. This condition is physically reasonable
and can be satisfied in widespread circumstances. Some examples are
invoked to illustrate that this condition can be realized by choosing
suitable spectrum distributions of oscillators making up the detector.

To quantitatively describe the intermediate state of decoherence between
the pure state and the most-largely mixed state, we need to calculate
the entropy of $S$
$$s=-\frac{K}{2}Tr(\hat{\rho}\ln\hat{\rho})\eqno(10)$$
for two cases: (I) For the two-level system and any finite $N$,
an exact solution for $s$ is obtained as the functional of $\hat{\rho}$.
(II) For the limits with very large and very small N,
the approximate solutions
of $s$ by certain perturbation methods. These calculations show that
the entropies indeed decrease as $N$ increases and they will take the maximum
values
at the infinite $N$. This means that an ideal macroscopic detector or
environment must cause the increment of entropy of the its monitored system .

The above general approach is used to built an exactly-solvable
dynamical model  for the quantum Zeno effect [25,26]
in connection with the recent experiment by Itano {\it et.al}[26]
about the inhibition of quantum transition between
the atomic energy levels. The present investigation compromises
the different points of view about this
 experiment testing quantum Zeno effect [27-37]. In this model
 the detector is simplified
as a system of $N$ oscillators with a suitable interaction with the measured
system--a two-level atom.   we show that, due to gradually-vanishing of the
off-diagonal elements in $\rho_S$, the two-level system will be frozen in
its initial level as the times of measurement in a given time
interval becomes infinite. This is just the quantum dynamical realization of
the QZE
through a  dynamical approach of the wavefunction collapse. The information
entropy for the process of quantum Zeno effect of two-level system is
calculated to manifest an interesting behavior of transition from random to
regularity:
for a given time interval, when the times $L$ of measurement is less than a
critical value $L_c$, the entropy changes at quite random as $L$ changes; when
$L$ is larger than $L_c$, the entropy decreases monotonically as $L$ becomes
larger.

 Finally, it   has to be pointed out that the correlation between the states
of the measured system and that of the detector has not been emphasized well
in the original HC model and its generalization . But
this problem was well analysed by Cini in his  beautiful dynamical model [39].
The present investigation is also to emphasize on both the
wavefunction collapse and
the state correlation in the Cini model. In fact, the correlation between the
states of
measured system and the detector is crucial for a realistic process of
measurement, which enjoys a scheme using the macroscopic counting number of
the measuring instrument-detector to manifest the microscopic state
of the measured system. Notice that
the original Cini model for the correlation between the states of measured
system S and the measuring instrument-detector D is build
only for a two-level
system interacting with the detector D, which consists of N
indistinguishable particle with two possible states $\omega_{0}$ and
$\omega_{1}$. For the two states $u_{+}$ and $u_{-}$, the detector has
different strengths of interaction with them.
Then, the large number N of "ionized" particle in the ionized state
$\omega_{1}$ transiting from the un-ionized state   $\omega_{0}$ shows
this correlations. In this paper, we wish to generalize the Cini's model
for the M-level system.\\

{\bf 2. Dynamical Description of the General Model } \\

In this section we wish to describe a general dynamical model for  quantum
decoherence caused by a suitable interaction between the measured system $S$
and a measuring instrument- detector (or an environment) $D$
that can be regarded as a reservoir at temperature $T$. The
considered system is only required to be of the non-degenerate
discrete spectrum. Let
$|n\rangle (n=1,2,...M)$  be the discrete eigenstates of
$S$ corresponding to $N$ energy levels
$E_n ~ (n=1,2,...M)$. Therefore, the Hamiltonian is formally  expressed as
$$
\hat{H}_s=\sum_{n=1}^ME_n|n\rangle{\langle}n|.
$$
$D$ is made up of $N$ particles with the single particle Hamiltonian
$\hat{H}_k(x_k)$ for  dynamical variables $x_k$
(such as  canonical coordinate
, momentum and  spin ) of  the k'th particle. Its Hamiltonian
$$\hat{H}_D=\sum^N_k\hat{H}_k(x_k)=\sum_{k,\alpha}\langle
\phi_{\alpha}|\hat{H}_k(x_k)|\phi_{\alpha}
\rangle|\phi_{\alpha}\rangle\langle\phi_{\alpha}| \eqno(2.2)$$
can be written in terms of $\hat{H}_k(x_k)$ or its eigenstates. Here,
it has been assumed
that there are not mutual interactions among detector's particles.

Physically,  the interaction between $S$ and $D$ can be
chosen to have the
different strengths for the  different states   of $S$. Thus, one can write
the interacting Hamiltonian  as
$$H_I=\sum_n\sum_kg(t)V_n(x_k)|n\rangle{\langle}n| \eqno(2.3)$$
where $g(t)$ ($=1$ for $0{\leq}t{\leq}\tau$; $=0$ for $t<0$ or $t>\tau$)
is a switching function; $V_n\neq V_m$ for $m\neq n$. Besides the
above-mentioned basic requirements, their is only a few of  constrains
on the model and one even need not to know the specific forms of both the
interaction and the Hamiltonians. Therefore,
it is reasonable to say our model is quite
universal in comparison with the previous models  for quantum measurement.

The system $S$ plus the detector  $D$ forms a composite system $C$ , the
``universe''.
In this sense, $C$ is closed and thus its evolution can be described by an
unitary operator $U(t)$ governed by the total Hamiltonian
$$H=H_s+H_D+H_I{\equiv}H_0+H_I$$
By changing into the interaction picture and then backing to the original
Schrodinger picture, a direct calculation gives the evolution operator
$$U(t)=U_0(t)U_I(t);$$
$$U_I(t)=\sum_nU_n(t)|n\rangle{\langle}n| \eqno(2.4)$$
where

$$U_0(t)=e^{\frac{1}{i\hbar}\hat{H}_0t}=\sum_{n=1}^M
e^{\frac{-iE_nt}{\hbar}}|n\rangle{\langle}n|\otimes
\prod_{k=1}^Ne^{\frac{-i\hat{H_k}(x_k)t}{\hbar}}$$
$$U_n(t)=\prod_{k=1}^NU_{n}^{[k]}(t)\eqno(2.5) $$
$$U^{[k]}_n(t)= \left\{ \begin{array}{ll}
         {\cal P}exp[\frac{1}{i\hbar}\int_{0}^{t}
         V_n[x_k(t^{'})]dt^{'}],& \mbox{for}~~ 0{\leq}t\leq\tau  \\
    U_n^{[k]}(\tau),& \mbox{for} ~~t\geq\tau
                \end{array}
     \right. $$
Notice that ${\cal P}$ is the time-order operator and
      $$x_k(t)=e^{\frac{-iH_kt}{\hbar}}x_ke^{\frac{iH_kt}{\hbar}}\eqno(2.6)$$
is the  representative of the variables $x_k$ of $D$ in
the interaction picture.

  It should be pointed out that the above mentioned evolution operator
$U_I(t)$ in the interaction picture just possesses the factorizable structure
found in ref.[14, 15] by this author, which may result in von Neumann's wave
packet
collapse in the quantum measurement.\\\\

{\bf 3. Quantum Decoherence at Finite Temperature}\\

If the system $S$  were closed, the Schrodinger time evolution should not
lead to the phenomenon of decoherence - wavefunction collapese
defined by eq.(1.1). In fact,
a pure state as a coherent superposition of some eigenstates of an observable
of $S$ can not evolve into a mixed state. This is because the unitary time
evolution operator $U(t)$ preserves the rank of density matrix for $S$,
which, however, has rank $1$ for a pure state and has rank larger than $1$
for a mixed state. We can also describe this impossibility in terms of the
definition of quantum entropy in  section  ?.

  However, in our present model, $S$ is considered as an open system
interacting with $D$.
Though the unitary evolution of the universe $C$ formed by $S$ plus $D$
can not
change $C$ to a mixed state from a pure state, its induced effective
evolution of $S$
given by removing the variables of $D$ may be non-unitary and thus the
rank of the reduced density matrix can be changed in the
evolution. In this sense, it is
possible to realize the decoherence  for quantum measurement
in a quantum dynamical process.

 Assume that  $D$ is in a mixed state, e.g., an thermal
equilibrium state at temperature $T$, when the interaction between $S$ and $D$
switches on.
This initial state of $D$ is denoted by the canonical distribution
      $$\rho_D(0)
      =\prod_{k=1}^{N}\otimes\sum_{n_k} P_{n_k} |n_k><n_k| \eqno{(3.1)}$$
where $ P_{n_k}$ is the classical probability of k'th particle of
the detector in the states $ |n_k>$
  $$ \sum_{n_k} P_{n_k} =1$$
 Let the initial state of $S$ is a pure
state, which is a coherent superposition
  $$|\phi >=\sum_{n=1}^{N}C_n|n> \eqno{(3.2)}$$
of the energy eigenstates of $S$ without the interaction with $D$. Then, we
write
down the density operator of the initial state for the universe $C=S+D$:
  $$\rho(0)=|\phi\rangle{\langle}\phi|\otimes\rho_D(0)={\sum}_{m,n}
  C_mC_n^*|m\rangle{\langle}n|\otimes\rho_D(0) \eqno{(3.3)}$$
Solving the von Neumann equation
  $$i\hbar\frac{\partial}{{\partial}t}\rho(t)=[\rho(t), \hat H] \eqno{(3.4)}$$
we obtain the density operator of $C$ at time $t$
  $$\rho(t)=U(t)^{\dagger}\rho(0)U(t)$$
Taking the partial trace of $\rho(t)$  over the variables of $D$, we have the
reduced
density matrix
  $$\rho_s(t)=Tr_D(\rho(t))=\sum_{n=1}^{M}|C_n|^2|n\rangle{\langle}n|
  +\sum_{n>n^{'}}^{M}(C_nC_{n'}^{*}F_{n,n}^{'}(N,t)|n\rangle
  {\langle}n^{'}|+h{\cdot}c) \eqno{(3.5)}$$
where the off-diagonal terms are accompanied by  factors
$$F_{n,n^{'}}(N,t)=e^{\frac{-i(E_n-E_{n^{'}})t}{\hbar}}Tr({U_{n^{'}}}^
{\dag}(t)U_n(t)\rho_D(0))$$
$$ = e^{\frac{-i(E_n-E_{n^{'}})t}{\hbar}}\prod_{k=1}^{N}\sum_{n_k}<n_k|
({U_n^{[k]}}^{\dag}(t)U_n^{[k]}(t)|n_k>P_{n_k}$$
$$= e^{\frac{-i(E_n-E_{n^{'}})t}{\hbar}}
\prod_{k=1}^{N}F_{n,n^{'}}^{[k]}(t) \eqno(3.6)$$
In the above expression, $Tr_k$ means taking partial trace
over the k'th variable
of $D$.

 Now, we are trying to find a central condition, under which as the
particle number $N$ of $D$ approaches  infinity, $F_{m,n}(N,T,t)$ approaches
zero for $m\neq n$,
$$
 |F_{m,n}(N,t)|\rightarrow 0 ~~as ~~N \rightarrow \infty   \eqno{(3.7)} $$
Eq.(3.7) determines the vanishing of the off-diagonal elements of $\rho_s(t)$
, that is to say,  it   will approximate the classical behavior
of the open system $S$ in
the macroscopic limit with very large $N$.

 To this end, we use the eigenstates $|\alpha\rangle$ of $H_k(x_k)$
 with the corresponding
eigenvalues $\varepsilon_{\alpha}(k)$ to rewrite the accompanying
factor $F_{n,n^{'}}(N,t)\equiv{F}$ . Then,
$$|F^{[k]}_{n,n^{'}}(t)|=
|\sum_{n_k}\langle n_k|U_{n^{'}}^{[k]\dag}(t)U_n^{[k]}(t)|
n_k{\rangle}P_{n_k}|$$
$$\leq\sum_{n_k}|\langle n_k|U_{n^{'}}^{[k]\dag}(t)U_n^{[k]}(t)|
n_k\rangle|P_{n_k}\leq\sum_{n_k}P_{n_k}=1\eqno(3.8)$$
Here, it has been taken into account that
$$|\langle n_k|U_{n'}^{[k]\dag}(t)U_n^{[k]}(t)|n_k\rangle|^2
=1-\sum_{m_k\neq n_k}|\langle m_k|U_{n^{'}}^{[k]\dag}(t)
U_n^{[k]}(t)|m_k\rangle|^2\leq 1,~~~ for~~t=\tau \eqno(3.9)$$
for an unitary operator $U_{n^{'}}^{[k]\dag} (t)U_{n}^{[k]}(t)$
and $n^{'}{\neq}n$.
In terms of the non-zero positive real number
     $$\Delta^{n,n'}_k(t)=-\lim|F_{n,n^{'}}^{[k]}(t)|\eqno(3.10)$$
the norm of accompanying factor $F_{n,n^{'}}(N,t)$ is  expressed as
$$ |F_{n,n^{'}}(T,N)|=e^{-\sum_{k=1}^{N}\Delta^{n,n'}_k} \eqno(3.11)$$
Due to  eq.(3.8), $\Delta^{m,n}_k\leq 0$; In general,
under a reasonable condition that there is at least one non-vanishing
off-diagonal element in a given row or  column  of each factor
$U_{n'}^{[k]\dag}(t)U_n^{[k]}(t) $, $\Delta^{m,n}_k(t)'$s
do not approach zero as
$k\rightarrow\infty$ at $t=\tau$ according to eq.(3.9). In this sense,
$ \sum_{k=1}^{\infty}\Delta^{m,n}_{k}(t)$  is
a diverging series
with the limit of infinity. In next section, some examples are present
to obey the above mentioned condition explicitly.

  The above discussion shows the possibility of realizing the
decoherence in a quantum dynamical  process at the macroscopic limit with
very large $N$. The above discussion does not depend on the specific forms
of both the single particle Hamiltonian $H_k(x_k)$ and the interaction
$V_n(x_k)$. It can be invoked to find a vast class of
quite general dynamical models for quantum measurement to describe quantum
decoherence - wave function collapse as the result of
the  dynamical evolution at macroscopic
limit .\\\\

{\bf 4. An concrete models with $D$ made up of oscillators}

\vspace{0.6cm}

In this section, we will use a typical model to explicitly illustrate how
the above approach works effectively for the dynamically-vanishing of the
off-diagonal elements of the evolving density matrix, concretely speaking,
under what kind of
circumstance the series
$\sum_{k=1}^{\infty}\Delta_k^{mn}(t)$ in the exponential
accompanying factor diverges into infinity. In this typical model,
$D$ is made up of $N$ oscillators with Hamiltonian
$$H=\sum_{i=1}^N\hbar\omega_ka^{+}_ka_k \eqno (4.1)$$
and the interaction between $D$ and $S$ is
$$H_I=\sum_{n=1}^M\sum_{i=1}^N\mu_ng_i(a_i^{+}+a_i)|n><n|\eqno (4.2)$$
where the requirement $\mu_m\neq \mu_n$ (for $m\neq n$) means that
the coupling of $D$ with $S$ has different strengths
for the different states $|n>$ of $S$.

In this sense, according to Wei-Norman's algebraic method [34],
a factor $U_n^{[k]}(t)$ of the evolution matrix can be assumed as
$$U_n^{[k]}(t)=e^{f_k^n(t)}e^{A_k^n(t)a_k^{+}}e^{B_k^n(t)a_k} \eqno (4.3)$$
Here, the coefficients $F_k^n(t), A_k^n(t)$, and $B_k^n(t)$ to be determined
satisfy a system of equations
$$\dot{A}_k^n=-\dot{B}_k^n(t)^{*}=\frac{g_k\mu_n}{i\hbar\omega}e^{i\omega t}$$
$$\dot{f}_k^n(t)=B_k^{n*}(t)\dot{A}_k^n(t) \eqno (4.4)$$
It leads to
$$A_k^n(t)=-B_k^{n*}(t)=\frac{\mu_ng_k}{\hbar\omega_k}[1-e^{i\omega_k t}]
\eqno(4.5)$$
and the real part of $f_k^n(t)$
$$Re(f_k^n(t))=-\frac{1}{2}|A_k^n(t)|^2=-\frac{\mu_ng_k}{\hbar\omega_k}
(1-\cos\omega_kt)\eqno (4.6)$$
is negative for $\omega_k\neq n\pi/t$.

To master the kernel of the problem, we consider an simple case with zero
temperature. The corresponding density matrix is $\rho_D(0)=|0><0|$.
Here, the ground state of $D$
$$|0>=|0_1>\otimes|0_2>\otimes...\otimes|0_N>$$
is a direct product of the vacuum states $|0_i>(i=1,2,\cdots,N)$ of $N$
oscillators. In this sense,
$$|F_{mn}^{[k]}(t)|=|<0|U_n^{[k]}(t)U_m^{[k]}(t)|0>|=e^{-\Delta_k^{mn}(t)}
\eqno (4.7)$$
where $$\Delta_k^{mn}(t)=2(\mu_m-\mu_n)^2\frac{g_k^2\sin^2\frac{\omega_kt}{2}}
{(\hbar\omega_k)^2}\equiv \eta_k(t))\eqno (4.8)$$

In the following discussion, we will detail the discussion about diverging
of the series $\sum_k\Delta_k^{mn}$ for various spectrum distributions of
$D$. The most simple case is that $D$ has a constant
discrete spectrum, i.e., $\omega_k=\omega=$constant,$ g_k=g=$constant.
In this case
$$|F_{mn}(N,t)|=exp\{-2N(\mu_n-\mu_n)^2\frac{g}{(\hbar\omega)^2}
\sin^2\frac{\omega t}{2}\}\eqno (4.9)$$
approaches zero as $N\longrightarrow\infty$ except for the period points
$\omega t=2k\pi (k=0,1,2...)$ . Generally, the series
$\sum_k\Delta_k^{mn}$ can
be repressed in terms of a un-specific spectrum distribution $\rho(\omega_k)$
as
$$ SR=2(\mu_m-\mu_n)^2\sum_k
\frac{\rho(\omega_k)g_k^2 \sin^2\frac{\omega_k t}{2}}
{(\hbar\omega_k)^2}\eqno(4.10)$$
In fact, in the case of discrete spectrum, the spectrum distribution means a
degeneracy that there are $\rho(\omega_k)$ oscillators possessing the same
frequency $\omega_k$. So long as the term
$$\rho(\omega_k)g_k^2 sin^2\frac{\omega_k t}{2}/(\hbar\omega_k)^2$$
does not approach zero as $k\longrightarrow \infty$,
the above series $SR$ must diverge  into infinity.
For example, if $\omega_k=k\omega$, $\rho(\omega_k)\propto k^{\eta+2}/g_k^2,
\eta>0$,then this series must diverge into infinity for each term
$\propto k^{\eta}\sin^2\frac{k\omega t}{2}>0$

Notice that $f(M,t)\equiv F_{m,n}(N,0,t)1$ and 0 means the complete coherence
and
the complete decoherence respectively.
 Because of the cut-off of frequency, the $\eta_k(t)$ do not approach zero
 except for the period points $ t\neq \frac{(2n+1)\pi}{\omega_k}$
 as $k\rightarrow \infty$.
Also due to $\eta_k>0$, the series $\sum_{k=1}^{\infty}\eta_k(t)$
must diverge into infinity, that is to say, $f(N,t)\rightarrow 0$ according
to eq.(3.4). Therefore, when the detector is macroscopic
$(N\rightarrow\infty),$
the off-diagonal elements in $\rho_S(t)$
vanishes and the wavefunction collapse appears as the result of the
dynamical evolution. It is realized from eq.(4.8) that, to realized the
wavefunction collapse,
we must constrict the interaction for each time measurement
only to take place in a time interval $\tau $ less than the oscillation
period $\tau_c=\frac{2\pi}{\omega_k}$. Otherwise, the coherence terms
suppressed
by the factor $f_k(t)\equiv F^k_{m,n}(0,t)$ 0will be resumed at the common
period points $t=t_c$,
 $\eta_k(t_c)=0$ and then $|f_k(t)|=1$. The figure 1 with a constant spectrum
 $\omega_k=\omega$ shows how
the decoherence appears in accompany with : $|f(N,t)|\rightarrow 0$ as $N$
increases for $t\neq t_c$, and how the coherence is resumed at $t=t_c$.
Figure 2 displays the same problems for the random spectrum distribution that
 the frequencies $\omega_k$ take random value with cut-off.

According the above analysis , one should let the interaction switch off
before the coherence restores in the wavefunction collapse quantum
dynamically. For this reason, the random spectral distribution and the
constant spectral distribution are much appreciated for our present study for
the QZE.
For the case with continuous spectrum, some interesting circumstances can
result from the concrete spectrum
distributions. In the first example with
$$\rho(\omega_k)=\frac{1}{g_k^2}$$
the series is convergence to a positive number proportional
to time t
$$SR=\sum_k\Delta_k^{mn}=\frac{2(\mu_m-\mu_n)^2}{\hbar^2}\int_0^{\infty}
\frac{sin^2\frac{\omega_kt}{2}}{\omega_k^2}d\omega_k=
\frac{\pi(\mu_m-\mu_n)^2t}{2\hbar^2}\eqno (4.11)$$
This shows that the norm of the accompanying factor $F_{mn}(0,t)$
is an exponential decaying factor, i.e,
$$|F_{m,n}(0,t)|=e^{-\frac{\pi(\mu_m-\mu_n)^2}{2\hbar^2}t}\eqno(4.12)$$
This quite interesting result is much similar to that in the quantum
dissipation [4].  As $t\longrightarrow\infty$ the off-diagonal elements of
density matrix vanish simultaneously!

In another example for continuous spectrum, the spectrum distribution is
Ohmic type (e.g,see ref.[4] ),i.e,
$$J(\omega)=\frac{\pi}{2}\sum_{j=1}^{N}\frac{g_j^2}{\omega_j}\delta
(\omega-\omega_j)=\eta\omega_j\eqno(4.13)$$
its alternative reformulation is
$$\rho(\omega_k)=\frac{2\eta\omega_k^2}{\pi g_k^2}\equiv\varepsilon
\frac{\omega_k^2}{g_k^2}\eqno (4.14)$$
Then
$$\sum_k\Delta_k^{mn}\longrightarrow
\frac{2(\mu_m-\mu_n)^2\varepsilon}{\hbar^2}\int_0^{\infty}
\sin^2\frac{\omega_k t}{2}d\omega_k=\infty ,for m\ne n\eqno (4.15)$$

In summary, so long as we choose a suitable spectrum
distribution of oscillators in the detector,
the series $\sum_k\Delta_k^{mn}$ can diverge into infinity, that is to say,
the dynamical evolution of $S$ plus $D$ can result in the wavefunction collapse
or quantum decoherence, though the discussion in this section was proceeded
with the oscillator detector, any detector (or environment)
weakly coupling to system may
be equivalent to a system of oscillators according to the proof given by
Caldeira and Leggett [4]. Therefore, the discussion in this section do not
lose the generality of the problem.\\\\

{\bf  5. Entropy Increment in Decoherence Process}\\

 Since quantum decoherence decreases the information available to
 the observation  about the
quantum open system $S$, the quantum entropy
$$S[\rho]=-\frac{K}{2}Tr({\rho}\ln\rho) \eqno{(5.1)}$$
as a functional of the density matrix $\rho$ can be used to characterize the
degree
of decreasing information quantitatively.

 In comparison with the statistical thermodynamics, the decoherence process
can be understood as an irreversible process in terms of the concept of
entropy. In fact, if $|\lambda\rangle$ is the eigenstate
of $\rho$ with eigenvalue $\lambda$, then
eq.(5.1) is re-expressed as
$$S[\rho]=-\frac{K}{2}\sum_{\lambda}~{\lambda}\ln\lambda \eqno{(5.2)}$$
Obviously, the entropy is invariant under an unitary transformation
and the time evolution of a closed system must not change its
entropy. In our problem, the entropies of the initial and final states are
zero and
$$S[\rho_{f}]=-\frac{K}{2}\sum_{n=1}^M|C_n|^2\ln|C_n|^2  \eqno (5.3 )$$
respectively,
manifesting that the decoherence process is certainly a process of increasing
entropy. Notice that eq.(5.2) defines the maximum entropy of the system $S$
for a given initial state (3.2) of $S$, which corresponds to complete
decoherence.

 In a well-established theory for quantum decoherence, it is expected that
the partially- and completely- decohering states are also described very well.
For this
reason, we calculate the corresponding entropy of the intermediate state
characterized by the reduced density matrix $\rho_s(t)$ for finite $N$.

 For the seek of simplicity, we first consider the two-state system with
$M=2$ physically. This case may be understood as spin $-\frac{1}{2}$
precession
interacting with a reservoir. In this sense, the reduced density matrix with
finite $N$ is explicitly written down
$$\rho_{s}=\left[\begin{array}{ll}
 |C_1|^2 & C_1C_2^*F \\
 C_2C_1^*F^* & |C_2|^2
   \end{array}
    \right] \eqno{(5.4)}$$
where
$$F=F_{12}(N,T)=e^{-\sum_{i=1}^{N}\Delta_{k}(t)}$$

In terms of its eigenvalues ${\lambda}_1=\frac{1}{2}(1+x)$ and
${\lambda}_2=\frac{1}{2}(1-x)$,
$$x=x(N)=\sqrt{1-4(1-|F|^2)|C_1|^2|C_2|^2}$$
$$=\sqrt{1-4(1-e^{-2\sum_{i=1}^{N}{\Delta}_k(t)})|C_1|^2|C_2|^2} \eqno(5.5)$$
the explicit expression of entropy is
$$S(N)=-\frac{K}{2}[(1+x)l_n(1+x)+(1-x)l_n(1-x)-2l_n2] \eqno{(5.6)}$$
where
$$||C_1|-|C_2|| \leq x\leq 1$$
    Since
$$\frac{dS(N)}{dx}=-\frac{K}{2}\ln\frac{(1+x)}{(1-x)}<0$$
and $x$ decreases as $N$ increases, $S=S(N)$ is a monotonically-increasing
function in the above domain of $x$.
When $N=\infty$, $x$ takes its maximum value so that there appears maximum
entropy
$$S_m=-\frac{K}{2}(|C_1|^2\ln |C_1|^2+|C_2|^2\ln|C_2|^2)\eqno (5.7)$$
The above  quantitative analysis shows us that the complete decoherence
means the maximum of entropy. For the sufficiently large $N$, the average
value $\Delta=N^{-1}\sum_{k=1}^{N}\Delta_k(t)$ is roughly independents for $N$
$$|F|=e^{-2N\Delta}$$
is approximated by exponential function of $N$.

 The explicit entropy functions for the partial decoherence in mesoscopic case
 with a finitely large N can  not
 be easily solved for a general system of $M(>2)$ energy levels.
 However, for few cases with very large $N$ or very small $N$,
 some approximation methods can be developed to calculate
 the entropy analytically.

Foe the case with small $N$ we define
  $$\Delta_{mn}(N)=F_{mn}(N)-1. \eqno(5.8)$$
Since the function $F_{n,n}=1$ and     $F_{m,n}(N)$  approaches the unity
for small $N$, the norm of
$\Delta_{m,n}(N)$  is quite small in this case. Thus, the reduced density (3.5)
can be decomposed into the
unperturbed part
$$\rho_0=\rho(0)={\sum}C_mC_n^*|m\rangle
{\langle}n|=|\psi\rangle\langle\psi|$$
and the perturbed part
$$\Delta\rho=\sum_{m,n=1}^M
\Delta_{m,n}(N)C_mC_n^*|m\rangle{\langle}n| \eqno{(5.9)}$$
The unperturbed part $\rho_{0}$ denotes a pure state and
then has M non-degenerate
eigenstates $|\psi\rangle$.
Choose $|\psi_0\rangle\equiv|\psi\rangle$ to be the first one of basis
vectors for the Hilbert space. Then,
the other N-1 basis vectors $|\psi_k\rangle=P|k\rangle~~~
(k=1,2,\ldots,N-1)$ to
$|\psi_0\rangle$ may be constructed in terms of the complementary projection
operator
$$P=1-|\Psi\rangle{\langle}\Psi|=\sum_{m=1}^{N}(\delta_{mn}-C_m^*C_n)
|n\rangle{\langle}m| \eqno{(5.10) }$$
It is not difficult to prove that such $N-1$ vectors are linearly-independent.
It follows from the Smite rule that N-1 vectors
$$|u_k\rangle=\sum_{k'}S_{k'k}|\psi_{k'}\rangle,~~k=1,2,...,N-1$$
are built as the orthogonomal basis for the complete space corresponding to
the zero eigenvalues of $\rho_0$. Therefore, the time independent
perturbation
theory determines the approximate eigenvalues of
$\rho=\rho_0+\Delta\rho$ up to of second order
$$\lambda_0{\simeq}1-\sum_{k=1}^{N-1}|\langle\Psi|\Delta\rho|u_k{\rangle}|^2
+\ldots \equiv 1-\delta+\ldots \eqno(5.11)$$
$$\lambda_k=|\langle\Psi|\Delta\rho|U_k\rangle|^2+\ldots \equiv
\delta_k+\ldots,k=1,2,\ldots M-1 $$
Obviously, the normalization of density matrix still holds
as $\sum_{k=0}^{N-1}\lambda_k=1$ under this perturbation. A direct
calculation from eq.(5.11)
results in the entropy function in small N limit as

$$S(N)= -\frac{k}{2}[(1-\delta)\ln(1-\delta)+\sum_k{\delta}_k\ln\sum_k
{\delta}_k]\eqno(5.12)$$
$$=-\frac{K}{2}\sum_{k=1}^{M-1}|\langle\Psi|\Delta\rho|u_k\rangle|^2$$

Another case that can be handled  analytically by using approximation method
is that  with
large N. In this sense, $|F_{mn}|$ is so small that the off- diagonal parts of
$\hat{\rho}$
(eq.(3.5)) can be regarded as a perturbation. The unperturbed part
$$\rho_M=\sum_{m=1}^M|C_m|^2|m\rangle{\langle}m|\eqno(5.13)$$
represents the mixed state with the  maximum entropy and
$|k\rangle$ and $|C_k|^2$ are just its eigenstates and
the corresponding eigenvalues. Invoking the time-independent
perturbation theory and regarding
$$\Delta \rho =\hat {\rho}-\rho_M\eqno(5.14)$$
as the perturbation, we write down the second's corrections for
the eigenvalue
$$\Delta\lambda_n^{[2]}=\sum_{m{\neq}n}
\frac{|F_{mn}|^2|C_m|^2|C_n|^2}{|C_m|^2-|C_n|^2}\eqno(5.15)$$
for the first order solution $\lambda_n^{(0)}=|C_N|^2$.
Then, the approximate entropy is obtained   up to of second order
$$S(N)=-\frac{K}{2}\sum_k[|C_k|^2(\ln|C_k|^2+2\Delta
\lambda_k^{(2)}\eqno(5.16)$$
\\\\

{\bf 6. Application to Quantum Zeno effect}\\

May years ago the quantum Zeno effect (QZE) was theoretically
proposed by Misra and Sudarshan [25] based on the postulate of wavefunction
collapse. It is argued that an unstable particle will never
be found to decay when it is continuously observed, more
generally speaking,
a frequent measurement inhibits the transitions between
quantum states. Recently, Itano, Heinzen, Bollinger and Wineland (IHBW)[26],
have reported that they have observed the QZE in an experiment about
atomic transition based on Cook's proposal. They claimed that the freezing
of  stimulated transition probability appeared when the two-level
atomic system is subjected to frequent measurements of the population of
a level. Then,  studies of the QZE have attracted much interest over
last years[27-37].

Among these discussions, Petrosky,Tusaki and Prigogin's (PTP's)
 remarkable work[27-28]
showed that the result in IHBW'S experiment can be recovered through
conventional quantum mechanics and do not involve a repeated collapse of
wavefunction. This is quite similar to Peres's observation[37] that the a
modified Hamiltonian may mimic the wavefunction collapse
to slow down the quantum transition and then realized the QZE in a pure
framework of quantum mechanics. It seems that the
conclusion of IHBW's experiment is challenged by the theoretical analysis of
PTP's work, and the same result can be obtained in term of both the use of
"wave-function collapse" by frequent measurement ,
or the use of the Schrodinger's
evolution as a pure quantum dynamical process[29,30]. Thus, It is difficult to
say whether the IHBW's experiment has proved the existence of the QZE or not.

To compromise the above mentioned different standpoints in a reasonable
framework, we will reconsider the QZE and its corresponding IHBW's experiment
from a distinct point of view.
Since the wavefunction collapse for quantum measurement can be regarded as a
quantum dynamical process in certain circumstance
according to the above discussion, it is natural to understand the QZE and
IHBW's experiment as the results of a measurement
monitoring the system continuously,
but the wavefunction collapse characterizing  measurement can be  realized
here quantum dynamically.\\

{\bf 6.1 Model Hamiltonian for QZE and its Evolution Matrix}\\

Now, we present a dynamical model for the QZE.
Let the measured object be a two-level systems $S$ with  Hamiltonian
$$
H_0=\frac{\Omega}{2}(|1><2|+|2><1|) \eqno(6.1)
$$
where $|1>$ and $|2>$ are the ground and excited states respectively.
$\Omega$ is  Rabi frequency of the external field coupling with
 atom. The continuous measurement for a given time interval $T$
is imaged as  a  limit of the $L$ times successive
 measurements at times $t=kT/L \ \ (0\leq k\leq L)$
with $N\longrightarrow \infty$ . For $k$'th measurement,
the interaction of  measuring
instrument--detector $D$ on the system turns
on at time $t_k=kT/L$ and then turns
off at
at time $t_k+\tau$ for $k=0,1,2,...$ .  Here, the  $\tau$ is very short
in comparison
with $T/L$. This case is quite similar to the  remarkable experiment by Itano
{\it et.al} [26]. In their experiment involving three levels of $ ^9Be^{+}$
ions,
a on-resonance radiation frequency field is applied to $ ^9Be^{+}$ with $L$
shot on-resonance optical pulses-measurement pulses to perform a quantum
measurement. Each optical pulse results in
the off-diagonal elements of the density
matrix to zero with the postulate of collapse wavefunction.

In our model, the instrument $D$  consists of $N$ harmonic
oscillators with free Hamiltonian
$$
H_D=\sum^{N}_{i=1}\hbar\omega_i a_i^{+}a_i\eqno(6.2)
$$
where $a_i^{+}$ and $a_i$ are  creation and annihilation operators
for $i$'s mode of boson states respectively. To realized
wavefunction collapse, the
frequency cut-off  must be introduced for the spectral distribution
of  the harmonic oscillators, that is to say, there exists a up-boundary
$\omega_c$ for the frequencies $\{ \omega_1,\omega_2,...,\omega_N\}$. The
model-interaction
$$
H_I=\frac{1}{2}\sum_{i=1}^{N}g(t)(a_i+a_i^{+})[(|1><1|-|2><2|)
\cos \frac{\Omega}{2}t\eqno(6.3)\\
 -\sin \frac{\Omega}{2}t(|2><1|-|1><2|)+1]
 $$

is turned-on and turned-off by the switching function
$$
g(t)=\left \{ \begin{array}{ll}
       g,&if~~~ t_k\leq t\leq t_k+\tau \\
       0,& otherwhere
\end{array}
\right.
$$                             \\

Based on the same  method as that in section 3,
we obtain the exact solution $U(t)$ for the
Schrodinger equation of the total system $C$ in terms of
the free evolution matrix
$$
U_0(t)=e^{-i\sum a_k^{+}a_k\omega_k t}\cdot e^{-\frac{iH_0}{\hbar}t}\equiv
U_D\cdot U_S\\
=(\prod_{k=1}^{N}e^{-ia_k^{+}a_k\omega_k t})\left[ \begin{array}{lr}
cos\frac{\Omega}{2}t & -isin\frac{\Omega}{2}t\\
-isin\frac{\Omega}{2}t & cos\frac{\Omega}{2}t
\end{array}
\right ],       \eqno(6.4)
$$
and the interacting evolution matrix
$$
U_e(t)=U_e(t)|1><1|+|2><2|\\
U_e(t)=\prod_{k=1}^M e^{F_k(t)}e^{A_k(t)a^{+}}e^{B_k(t)a}\equiv
\prod_{k=1}^M U^{[k]}(t) \eqno(6.5)
$$
where

$$
A_k(t)=-B^*_k(t)=\int_0^t\frac{g(t)}{i\hbar}e^{i\omega_k t}dt=\frac{g}{i\hbar}
[\sum_{j=0}^{l-1}\int_{t_j}^{t_j+\tau}e^{i\omega_k t}dt +\int_{t_l}^{t_l+t}
e^{i\omega_kt}dt]$$
$$=\frac{g}{\hbar \omega}[(1-e^{i\omega_kt})e^{\frac{i\omega_klT}{N}}
+(1-e^{i\omega_k\tau})
\frac{1-e^{\frac{i\omega T l}{L}}}{1-e^{\frac{i\omega T}{L}}}],\eqno(6.6)$$
$$\dot{F}_k(t)=B^*_k(t) \dot{A}_k(t) \eqno(6.7)
$$
for $ t_l \leq t \leq t_l+\tau $. Here, we have used the properties of
the switching function $g(t)$. If $\tau <<1$, then the approximate
function
$A_k(t)$ is
\begin{equation}
A_k(t)=\frac{g}{\hbar\omega}(1-e^{i\omega_k t})e^{\frac{i\omega l\tau}{L}}
\end{equation}
It follows from the above equations that
$$\eta_k(t))
Re[F_k(t)]=-\frac{1}{2}\int_0^t(\dot{A}_k(t^{'})A_k^{*'}(t)+\dot{A}_k(t^{'})^{*}
A_k(t^{'}))dt=-\frac{1}{2}|A_k(t)|^2 \eqno(6.9)
$$
Notice that real part of $F_k(t)$ is negative and $e^{Re(F_k(t))}$
is not larger than unity.  \\\\

{\bf 6.2 QZE based on Dynamical Collapse}    \\

For the measured system $S$ with an initial pure  state
$$
|\psi>=c_1|1>+c_2|2>
$$
and the measuring instrument $D$ with initial mixture state described by a
density matrix
$$\rho_D=\prod_{k=1}^N\rho_D(k),$$
the initial state of
the composite system $C$ formed by $S$ plus $D$ is a mixture state
with density matrix
$$
\rho(0)=|\psi><\psi|\otimes \rho_D
$$
Then, the state of $C$  at $t(<t_1)$ is
$$
\rho(t)=U_0(t)U_e(t)\rho_{(0)}U_e^{+}(t)U_0^{+}(t)
$$

Now, we understand the process of measurement as a procedure to determine
 the reduced density matrix,
 which contains the total information of the measured system $S$.
The first measurement results in a reduced density matrix for $S$
$$
\rho_s=Tr_D\rho(t)=U_S(t)(|c_1|^2|1><1|+|c_2|^2|2><2|+\\
f(N,t)|1><2|+f(N,t)^{*}|2><1|)U_S^{+}(t) \eqno(6.10)
$$
 by taking the trace for the variables of the detector $D$ where the
 coherence factor
$$
f(N,t)=Tr_D(U_e(t)\rho_D)=\prod_{k=1}^{N}f_k(N,t)\equiv\prod_{k=1}^N
Tr_k[U^{[k]}(t)\rho_D(k)]
$$
is factorizable. This factorization is crucial for the appearance
of wavefunction collapse . In terms of
the probabilities $P_n(k)$ of its distribution on
the Fock state $|n>$ for the k'th oscillator in  measuring instrument,
$\rho_D(k)$ can be re-expressed as
$$
\rho_D(k)=\sum P_n(k)|n><n|
$$
we prove
$$
|f_k(t)|=|Tr_k(U^{[k]}(t)\rho_D(k))|=
|\sum P_n(k)<n|U^{[k]}(t)|n>|\nonumber\\
        \leq\sum P_n(k)|<n|U^{(k)}(t)|n>|\leq\sum P_n(k)=1
$$
with a similar discussion to that in section 3.
Thus, the $N$- multiple product of all $|f_k(t)|$'s
must approach zero as $N\longrightarrow\infty$ unless most of $|f_k(t)|$'s
are
unity simultaneously. It will be illustrated by the following typical
example that the case that all or most of $|f_k(t)|$'s are unity is
rather special and this case can be eliminated by choosing a spectrum
distribution with cut-off of frequency.
In this example, we take $\rho_D(k)$ to a pure state $|0><0|$,obtaining
$$
f(N,t)=<0|U_I(t)|0>=e^{-\sum_{k=1}^{N}\eta_k(t)}\eqno(6.12)
$$
where $|0>=|0_1>\bigotimes |0_2>\bigotimes...\bigotimes|0_M>$
is the vacuum state of the detector $D$ consisting of $M$ oscillators.
$$
\eta_k(t)=-Re[F_k(t)]=\frac{1}{2}|A_k|^2=
\frac{g^2}{(\hbar\omega_k)^2}(1-cos\omega_k t)\eqno(6.13)
$$

Notice that eq.(6.13) is quite similar to eq.(4.8).
Then,, when the detector is macroscopic
$(N\rightarrow\infty),$
the off-diagonal elements in $\rho_S(t)$
vanishes and the wavefunction collapse appears as the result of the
dynamical evolution, that is
$$
\rho(t)\longrightarrow\rho_S=U_S(t)(|c|^2|1><1|+|c_2|^2|2><2|)
U^{+}_S(t)\eqno(6.14)
$$

In the following discussion,
we use the above developed theory for the quantum
dynamical model of wavefunction collapse to recover the QZE.

According to eq.(3.4), after the first time of  measurement, the density
matrix is
$$
\rho_s(t_1=\frac{T}{L})=(c_1^{[1]}|1><1|+c_2^{[2]}|2><2|+A^{[1]}|1><2|+A^{[1]*}
|2><1|)+O^{[1]}(\frac{1}{N})\eqno(6.15)
$$
where
$$c_1^{[1]}=|c_1|^2\alpha^2+|c_2|^2\beta^2\\
c_2^{[1]}=|c_2|^2\alpha^2+|c_1|^2\beta^2\\
A^{[1]}=\frac{i}{2}|c_1|^2\beta-\frac{i}{2}|c_2|^2\eqno(6.16)$$
$$
\alpha=\alpha(L)=cos\frac{\Omega T}{2L},\ \
 \beta=\beta(L)=sin\frac{\Omega T}{2L}
$$
Notice that the last term in eq.(4.1) is
$$
O^{[1]}_K(\frac{1}{N})=U_S^{(t)}[f(N,t)|1><2|+
f^{*}(N,t)|2><1|]U^{+}_S(t) \eqno(6.17)
$$
which vanishes as $N\longrightarrow \infty$, the first term in eq.(3.1)
$$
U_S(t_1)(|c_1|^2|1><1|+|c_2|^2|1><1|)U^{+}_S(t_1)
$$
represents the mixed state with complete decoherence.
Subsequently, the measurement of the first time cancels the off-diagonal term
of
$\rho_s(0)=|\psi><\psi|$ as $N\longrightarrow\infty.$
 The measurement of
second time will cancel the off-diagonal terms $A^{[1]}|1><2|$
and $A^{[1]*}|2><1|$ in eq(3.1).
Similarly, after the measurement of $k$'th time
$$
\rho_s[k]=|c_1^{[k]}|^2|1><1|+
\  |c_2^{[k]}|^2|2><2|
+A^{[k]}|1><2|+A^{[k]*}|2><1|+O_k^{[1]}(\frac{1}{N})$$
$$
=U_S(t_k)(|c_1^{[k-1]}|^2|1><1|+|c_2^{[k-1]}|^2|2><2|)
U_S(t_k)^{+}+O_k^{[1]}(\frac{1}{M})\eqno(6.18)
$$
where $c_1^{[k]}$ and $c_2^{[k]}$ are determined by the recurrent relations
$$
c_1^{[k]}=c_1^{[k-1]}\alpha^2+c_2^{[k-1]}\beta^2,\\
c_2^{[k]}=c_2^{[k-1]}\alpha^2+c_1^{[k-1]}\beta^2 \eqno(6.19)
$$
where
$$O_k^{[1]}(\frac{1}{N})=U_s(t_{k-1})[A^{[k-1]}f(N,t_k)|1><2|+A^{[k-1]*}
f(N,t_k)^{*}|2><1|]U_s^{+}(t_{k-1})$$
also disappear as $N\longrightarrow\infty $

It is quite difficult to solve $c_1^{[k]}$ and $c_2^{[k]}$ explicitly
from eq. (6.19), but we can invoke the computer simulation to
evaluate the variations of $c_1^{[L]}$ and $c_2^{[L]}$ as the
measurement times $L$
increase. If the system is initially in pure state $|1>$,
the initial conditions are $c_1=1$ and $c_2=0$. It is illustrated in Figure 3
 that , as $L\longrightarrow \infty$, the distributions
 $|c_1^{[L]}|\longrightarrow \infty$ while $|c_2^{[L]}|\longrightarrow 0.$
 Notice that $|c_1^{[L]}|$ first decrease and then increase to 1. This
 means that, for a macroscopic instrument ($N\longrightarrow\infty$),
 the system will be forced back to  the state $|1>$ as the successive
 measurements become continuous for $L\longrightarrow\infty$. This
 just the QZE! \\

{\bf 6.3 ``Transition" of Quantum Entropy for QZE}\\

Now, we consider the entropy of the measured system $S$ as a
function of the times $N$ of measurement.
Considering that the entropy is invariant under an unitary transformation due
to the original definition , one can obtain the entropy for the QZE after
$L$'th  measurement
$$
S(L)=S[\rho_S(L)]=-\frac{k}{2}(|c_1^{[L]}|\ln|c_1^{[L]}|+
|c_2^{[L]}|\ln|c_2^{[L]}|)\eqno(6.20)
$$
The entropy $S=S[\rho_S(L+1)]$  takes its maximum value
$$
S_{max}(L+1)=\frac{k}{2}\ln 2\eqno(6.21)
$$
only when
$$
|c_1^{[L]}|=|c_2^{[L]}|.
$$
According to the recurrent relations eq.(6.19), the above equation leads to
$$
|c_1^{[L]}|-|c_2^{[L]}|=(\alpha^2-\beta^2)(|c_1^{[L-1]}|-|c_2^{[L-1]}|)
=(\alpha^2-\beta^2)^L(c_1^2-c_2^2)
$$
Then, it is easily observed  that
the above equation holds only when
$$\alpha^2(L)=\beta(L)^2$$ or
$$
\frac{\omega T}{L}=\frac{\pi}{2}(2l+1), l=0,\pm 1,\pm 2,...\eqno(6.22)
$$
The entropy $S$ takes several maximum values
$$
S_m(L)=-\frac{k}{2}[c_1^{[\bar{L}]}\ln c_1^{[\bar{L}]}
+c_2^{[\bar{L}]}\ln c_2^{[\bar{L}]}] \eqno(6.23)
$$
where
$$
\bar{L}=[\frac{2\omega T}{(2l+1)\pi}],\ \ \ l=0,\pm 1,\pm 2,.... \eqno(6.24)
$$
Here, $[x]$ denotes the integer part of the real number for $x\geq 0$, if
$x\leq 0$ then $ [x]=0.$

Let us consider some special examples
for the above discussions.
If $\omega T=\pi, \bar{L}=2$ ( for $l=0)$,
and there only one point for the maximum entropy
$$
S_{L+1}=-\frac{k}{2}(|c_1^{[2]}|^2\ln|c_1^{[2]}|^2
+|c_2^{[L]}|^2\ln|c_2^{[L]}|^2)\eqno(6.25)
$$
Generally, if $\omega T=k \pi $ for positives integer $k$ ,
$$
\bar{L}=[\frac{2k}{2l+1}]=2k,[\frac{2k}{3}],[\frac{2k}{5}],...,
[\frac{2k}{2X+1}]\eqno(6.26)
$$
where $X\leq k-1,$ that is to say,
there are $k$ points $2k,[\frac{2k}{3}],[\frac{2k}{5}],...,[\frac{2k}{2k+1}]$
for the maximum values of entropy.

For the general case with $\omega T=k T$, the above analysis shows us that,
if $L$ is less than the critical value $L_c=2k$, the variation of $S(L)$
is a "random function" of $L$, which is not monotonic; however,when $L$ is
larger than the critical value $2k$, $S(L)$ is a
monotonically-decreasing function of $L$ . When $\omega T$ is not an
integer times of $\pi$ ,the critical point for $\bar{L}$ is
$\frac{2\omega T}{(2l+1)}\pi$. Such a feature of "transition" from random to
regularity for quantum entropy in the QZE is illustrated in Figures 4 and 5.
Physically, the QZE defines a transition of the information entropy
from random to regularity.
\vspace{1cm}

{\bf 7 Generalization of Cini's Model for Quantum Measurement}\\

        It has to be pointed out that the correlation between the states
of the measured system and that of the detector has not been emphasized well
in the original Hepp-Coleman model and its generalization .
This problem was well analysed by Cini with a beautiful dynamical model [39].
The present investigation is  to emphasize on both the
wavefunction collapse and
the correlation collapse for a generalization of the Cini model.
In fact, the correlation between the states of
measured system and the detector is crucial for a realistic process of
measurement,which enjoys a scheme using the macroscopic counting number of
the measuring instrument-detector to manifest the microscopic state
of the measured system.

The original Cini model for the correlation between the states of measured
system S and the measuring instrument-detector D is build for a two-level
system interacting with the detector D, which consists of
indistinguishable particle with two possible states $\omega_{0}$ and
$\omega_{1}$ . For the two states $u_{+}$ and $u_{-}$, the detector has
different strengths of interaction with them.
Then, the large number N of "ionized" particle in the ionized state
$\omega_{1}$ transiting from the un-ionized state   $\omega_{0}$ shows
this correlations. \\

{\bf 7.1 Generalized Cini Model}\\

In this section, we wish to generalize the Cini's model
for the M-level system. The measured system S with M-levels has the model
Hamiltonian
$${\hat H}_{S}=\sum_{k=1}^{M}E_{k}|\Phi_{k}><\Phi_{k}|\eqno{(7.1)}$$
Where $|\Phi_{k}>$  are the eigenstates corresponding to the eigenvalues
$E_{k}$ $(k=1,2,\cdots,M)$. The detector D is a two-boson-state system with
the free Hamiltonian
$$\hat H_D=\hbar\omega_1a_1^+a_1+
\hbar\omega_2a_2^+a_2\eqno(7.2)$$
where $a_{i}$ and $a_{i}^{+}$ are the creation and annihilation operators
and they satisfy
$$[a_{i},a_{j}^{+}]=\delta_{ij}  ,
\;\;   [a_{i},a_{j}]=[a_{i}^{+},a_{j}^{+}]=0\eqno(7.3)$$
In the $Schr\ddot odinger$ representation, the interaction is described by
$$\hat H_{I}(t)=\sum_{n}ge^{-\eta t}(w_{n}|\Phi_{n}><\Phi_{n}|)
(e^{i(\omega_{2}-\omega_{1})t}a_{1}^{+}a_{2}+e^{i(\omega_{1}-\omega_{2})t}
a_{2}^{+}a_{1})\eqno(7.4)$$
where the non-degenerate weights $w_{n}$ represent the different
strengths for the different states $|\Phi_{n}>$ of the system. The exponential
decay factor $e^{-\eta t}$ for $\eta>0$ is here introduced to turn off the
interaction after suitable time so that the coherence can not restore in the
evolution process. This point can be explicitly seen in the following
discussion. The introduction of time-dependent factors
$e^{\pm i(\omega_{1}-\omega_{2})t}$  is
quite similar to that in ref.[13] where these factors are used to describe the
energy exchange due to the presence of the free Hamiltonian $H_{D}$.
Notice that there was not the free Hamiltonian for detector in the original
Hepp-Coleman model such as $H_{D}$ in our present model.
Let $H=H_{S}+H_{D}+H_{I}$   be the total Hamiltonian in the
$Schr\ddot odinger$ representation for the composite system formed by S
plus D. Transforming the problem into the interaction representation
with the evolution operator
$$U_{0}{(t)}=exp[\frac{1}{i\hbar}(H_{S}+H_{D})t]$$
one has the interaction potential
$$V_{I}(t)=e^{-\eta t}g\sum_{n=1}^{M}W_{n}|\Phi_{n}><\Phi_{n}|(a_{1}^{+}a_{2}
+a_{1}a_{2}^{+})\eqno(7.5)$$
In order to diagonalize $V_{I}(t)$, we invoke the canonical transformation as
in ref.[5]
$$a_{1}=\frac{1}{\sqrt2}(b_{1}-b_{2}),\;
a_{2}=\frac{1}{\sqrt2}(b_{1}+b_{2})\eqno(7.6)$$
where the new boson operators $b_{i}$ and $b_{i}^{+}$ satisfy the same
bosonic commutation relation. In terms of these operators, $V_{I}(t)$ is
rewritten as the diagonal form
$$V_{I}{(t)}=ge^{-\eta t}\sum_{n=1}^{M}W_{n}|\Phi_{n}><\Phi_{n}|(b_{1}^{+}b_{1}
-b_{2}^{+}b_{2})\eqno{(7.7)}$$
Then, considering that the interaction part $V_{I}(t)$ commutes with each
other at different time, ie.
$$[V_{I}(t),V_{I}(t')]=0$$
one can express the evolution operator
$$U_{I}(t)=exp[\frac{1}{i\hbar}\int_{0}^{t}{V_{I}(t')dt'}]=\sum_{n=1}^{M}
exp[-it_{\eta}gW_{n}(b_{1}^{+}b_{1}-b_{2}^{+}b_{2})]|\Phi_{n}><\Phi_{n}|$$
$$\equiv\sum_{n=1}^{M}U_{n}(t)|\Phi_{n}><\Phi_{n}|\eqno(7.8)$$
where
   $$t_{\eta}=\frac{1-e^{-\eta t}}{\eta}$$
takes the real time as $\eta\rightarrow 0$.
It can be regarded as the $\eta$-
deformation of time t; when $t\rightarrow\infty$,
$t_{\eta}\rightarrow\frac{1}{\eta}$.         \\

{\bf 7.2  Correlation of states from Evolution of State}\\

Now, we consider the evolution of the total system starting with an initial
state at t=0
$$|\Psi(0)>=\sum_{k=1}^{M}C_{k}|\Phi_{k}>\otimes|N,0>\eqno(7.9)$$
where
$$|m,n>=\frac{a_{1}^{+m}a_{2}^{+n}}{\sqrt{m!n!}}|0>\eqno(7.10)$$
denotes a Fock state of two-boson system
which denotes that there are n particles in "ionized" state and m particles
in un-ionized state. It is hoped
to manifest a correlation between the states $|m,n>$
of detector and the state $|\Phi_{l}>$ of the system in a dynamical
evolution of the state $|\Psi(t)>$ in some limiting case so that one can
read out the state $|\Phi_{l}>$  from the manifestation of the state
$|m,n>$ of the detector.
Notice that the eigenstates of the operator
$$\hat O=b_{1}^{+}b_{1}-b_{2}^{+}b_{2}\eqno(7.11)$$
are
$$|\lambda,N-\lambda \}=\frac{1}{\sqrt{\lambda!(N-\lambda)!}}b_{1}^{+\lambda}
b_{2}^{+(N-\lambda)}|0>$$
$$={(\frac{1}{\sqrt{2}})}^{N}\sum_{m=1}^{\lambda}\sum_{n=1}^{N-\lambda}
\frac{\sqrt{\lambda!(N-\lambda)!}}{\sqrt{(\lambda-m)!(m!)(N-\lambda-n)!(n!)}}
|m,n>\eqno(7.12)$$
with the eigenvalues
$$\varepsilon_{N}(\lambda)=2\lambda-N\eqno(7.13)$$
where $\lambda=0,1,2,\cdots,N$. for a given integer $N$. The original Fock
state can be expended in terms of $|\lambda,N-\lambda>$ as
$$|N,0>={(\frac{1}{\sqrt{2}})}^{N}\sum_{\lambda
=0}^{N}\frac{\sqrt{N!}(-1)^{N-\lambda}}
{\sqrt{\lambda!(N-\lambda)!}}|\lambda,N-\lambda\}\eqno(7.14)$$
Then, one obtains the wavefunction at time t
$$|\Psi_{I}(t)>=U_{I}(t)|\Psi(0)>$$
$$=\sum_{\lambda=0}^{N}\sum_{k=1}^{M}C_{k}({\frac{1}{\sqrt{2}}})^{N}$$
$$\frac{\sqrt{N!}(-1)^{N-\lambda}}{\sqrt{\lambda!(N-\lambda)!}}
e^{-igt_{\eta}W_{k}(2\lambda-N)}|\Phi_{k}>\otimes|\lambda,N-\lambda>
\eqno(7.15)$$
Then, one has
$$|\Psi_{I}(t)=\sum_{k=1}^{M}C_{k}|\Phi_{k}>\otimes\sum_{n=0}^{N}
a_{n}(t,k)|n,N-n>\eqno(7.16)$$
where
$$a_{n}(t,k)=
\frac{(-1)^{N-n}\sqrt{N!}}{\sqrt{n!(N-n)!}}
cos^{n}(gW_{k}t_{\eta})sin^{N-n}(gW_{k}t_{\eta})\eqno(7.17)$$
Obviously, the probability
of finding n ``ionized" particles in second bosonic mode is
$$P_{n}=|a_{n}|^{2}=\frac{N!}{n!(N-n)!}cos^{2n}(gW_{k}
t_{\eta})sin^{2(N-n)}(gW_{k}t_{\eta})\eqno(7.18)$$
or
$$P_{n}=C_{n}^{N}p^{n}(1-p)^{N-n}\eqno(7.19)$$
where
$$C_{n}^{N}=\frac{N!}{(N-n)!n!} , \; p_{k}(t)=cos^{2}(gW_{k}t_{\eta})$$
when N is very large so that the Stirling formula is valid, it can be proved
that when $n_{k}=\bar n_{k}=Np_{k}(t)$,
the probability $p_{n}$ has its maximum.
$$P_{\bar n_{k}}=C_{\bar n_{k}}^{N}(\frac{\bar n_{k}}{N})(\frac{N-\bar n_{k}}
{N})^{N-\bar n_{k}}\eqno(7.20)$$
Notice that the derivation is the same as that in ref.[39], but $\bar n_{k}$
depends on the index k. As proved in ref.[39], $P_{n_{k}}$ is very strongly
peaked around its maximum $P_{\bar n_{k}}$, which becomes unity
when $N\rightarrow\infty$. Therefore, if the detector is very
macroscopic ($N\rightarrow\infty$), then
$P_{n_{k}}(N\rightarrow\infty)=\delta_{n_{k}\bar n_{k}}$
that leads to
$$|\Psi_{I}(t)>_{N\rightarrow\infty}=
\sum_{k=1}^{M}C_{k}|\Phi_{k}>\otimes|\bar n_{k}(t),N-\bar
n_{k}(t)>\eqno(7-21)$$
When $\bar n_{k}(t)\ne\bar n_{k'}(t)$ for $k\ne k'$,
a one-to-one correlation between the states $|\Phi_{k}>$ of S and the states
of D. In this sense, if the detector is found in the state
$|\bar n_{k}(t), N-\bar n_{k}(t)>$,
it can be concluded that the system is in this state $|\Phi_{k}>$
   A realistic detector must have a good fringe visibility, which can
manifests the macroscopic differences between any two states of
$|\bar n_{k}, N-\bar n_{k}>$ for different k.
It required that there is not the considerable overlap between
$|\bar n_{k}, N-\bar n_{k}>$ and $|\bar n_{k'}, N-\bar n_{k'}>$ for $k\ne k'$.
In fact, if $\bar n_{k}=\bar n_{k'}$
Then
$$gW_{k}t_{\eta}=gW_{k'}t_{\eta}+n\pi, n=0,1,2,\cdots $$
or
$$gt_{\eta}(W_{k}-W_{k'})=n\pi$$

   Otherwise, if one lets the interaction between S and D decay very fast so
that the limiting time
$$\frac{1}{\eta}\le\frac{\pi}{g(W_{k}-W_{k'})}$$
for any $k\ne k'$, then $\bar n_{k}(t)\ne\bar n_{k'}(t)$
for any long time evolution.
Notice that the above mentioned problem of overlap
of correlation states is a disadvantage in the original Cini's model, where
there is not the decay factor of interaction. So, in the time
$t_{n}=\frac{n\pi}{gW}$
the states $|\bar n, N-\bar n>$ and $|N, 0>$
are completely overlap. Here, it is the case of two levels  with
$W_{2}=W=\frac{1}{\sqrt{N}}, W_{1}=0$, thus, at $t=t_{n}$
the correlations vanish for the original Cini's model. Introducing the decay
factor $e^{-\eta t}$ in the interaction is the key point to avoid vanishing
of correlation in our model.
     In fact, such decay of interaction can appear in realistic physics.
For example, an atom is prepared in a microwave cavity loaded with an
electromagnetic field which can decay at a suitable rate.
In this example,
the atom and the cavity are regarded as the system and the detector
respectively. Notice that this example is quite useful for the studies of
atomic cooling [40].\\

{\bf 7.3 Wavefunction Collapse in Cini Model}\\

  We can also use the above generalized Cini's model to
describe the wavefunction collapse for the M-level system
quantum mechanically.
  Let the system D be initially prepared in a coherent superposition of
M-level
$$|\Psi>=\sum_{k=1}^{M}C_{k}|\Phi_{k}>\eqno{(7.22)}$$
and the detector be adjusted in the initial state $|N,0>$,
then the density matrix for the initial state of total system is expressed as
$$\hat\rho_{0}=\sum_{k, k'}C_{k}C_{k'}^{*}
|\Phi_{k}><\Phi_{k'}|\otimes|N,0><N,0|\eqno{(7.23)}$$
Using the evolution operator $U_{I}{(t)}$ in the interaction representation,
one formally write down the density matrix for the total system at t
$$\hat\rho(t)=U_{I}(t)\rho_(0)U^{+}_{I}(t)$$
Because we are only interested in the final state of the system other than
that of the detector for the consideration of wavefunction collapse, so
we must take trace for the variable of detector in the
total density matrix to obtain a reduced density matrix for S
$$\hat\rho_{S}(t)=Tr_{D}\rho(t)=\sum_{m,m'}<m,m'|\rho(t)|m,m'>$$
$$=\sum_{k}|C_{k}|^{2}|\Phi_{k}><\Phi_{k'}|$$
$$+\sum_{k\ne k'}e^{-it_{\eta}(E_{k}-E_{k'})/\hbar}
cos^{N}[g(W_{k}-W_{k'})t_{\eta}]C_{k}C_{k'}^{*}|\Phi_{k}><\Phi_{k'}|
\eqno(7.24)$$
where we have used
$$\sum_{n=0}^{N}|a_{n}(t,k)|^{2}=1\eqno(7.25)$$
and
$$\sum_{n=0}^{N}a_{n}(t,k)a_{n}^{*}(t,k')
=cos^{N}[g(W_{k}-W_{k'})t_{\eta}]\eqno(7.26)$$
Notice that each off-diagonal element in the density matrix is accompanied
by a time-dependent factor
$$F^{N}(k,k')=cos^{N}[g(W_{k}-W_{k'})t_{\eta}]\eqno(7.27)$$
which is a N-multiple product of factors $cos[g(W_{k}-W_{k'})t_{\eta}]$.
Recalling that due to the existence of the strong decay factor $e^{-\eta t}$
so that
$$\eta\ge\frac{g(W_{k}-W_{k'})}{\pi}$$
 holds for any $k\ne k'$,
we observed that the deformed time $t_{\eta}$ changes from $t_{\eta}=0$ to
$t_{\eta}=\frac{1}{\eta}$ as the real time changes from $t=0$ to
$t\longrightarrow\infty$ respectively. In this sense,
$$g(W_{k}-W_{k'})t_{\eta}<\pi\eqno(7.28)$$
and then
$$0\le |cos[g(W_{k}-W_{k'})t_{\eta}]|<1$$
or
$$|cos[g(W_{k}-W_{k'})t_{\eta}]|=
e^{-f_{k}(t_{\eta})},f_{k}(t_{\eta})>0\eqno(7.29)$$
This observation leads to
$$F^{N}(k,k')=e^{-Nf_{k}(t_{\eta})}$$
which obviously approach zero as $N\rightarrow\infty$.
Therefore, when the detector is macroscopic, $(N\rightarrow\infty)$
the off-diagonal terms of the density matrix vanish and the wavefunction
collapse is realized quantum dynamically.\\

  To end this paper, we present some comments on the above discussions.
Though this paper provides  one with an extensive generalization
and the unified description for a number of dynamical models (
e.g., the HC model ) for the
quantum decoherence, we have to say that a disadvantage in the
original model  still exists in the present models. This is the
oscillation of ${U_n^{[k]}(t)}^{\dagger}U_{n'}^{[k]}$ may enable
most of the
factors $F_{n,n'}^{[k]}(T,t)$ in eq.(3.6) to become unity  at a specific
time
$t=\tau_0$ and thus the whole accompanying factor can not approach zero
at this time.
To suppress such kind of oscillation so that
the decoherence appears in dynamical
evolution, a phenomenological method is to use the switching function
$g(t)$ in the interaction (2.3). However, the
microscopic mechanism of this switching
is not clear for us. We believe that the quantum dissipation caused by
the detector or environment
is a possible way to introduce such a switching mechanism microscopically.
For the concrete example that $D$ is made up of harmonic oscillators, the
discussion in section 4 showed that this kind of dissipation may result
from the specific distribution. How to realize the  quantum decoherence
directly through quantum dissipation is still an open question.
\vspace{1cm}

{\bf Acknowledgements}

The author gratefully acknowledges the support of K.C.Wong Education Foundation
Hong Kong.
The author thanks Professor C.N.Yang
for drawing his attentions to the recent progress in the quantum measurement
theory and the phenomenon of quantum dissipation  with decoherence.
He also thank Professor Tso-hsiu Ho for helpful discussions.

\newpage
\begin{center}
{\large \bf Figure Captions}
\end{center}
\ \ \\
Figure 1: The norm $|f(N,t)|$ of coherence factor as the function of time $t$
for the case with spectral
distribution $\omega_k=\omega$. Figures 1-a, 1-b, and 1-c correspond to
the times of measurement equal to 5,10 and 500 respectively. [Here
, $\omega_k=\omega=0.1, \frac{g}{(\hbar\omega_k)^2}=0.00001$]. For larger $N$
(e.g. $N$=500), the coherence enjoyed by $|f(N,t)|$ almost disappear for
$\bar{t}_k<t<\bar{t}_{k+1}$. However, at $t_k=\frac{2\pi k}{\omega},
(k=1,2,...,N)$ the coherence is resumed in a very shot time.
\ \ \\
\ \ \\
Figure 2. The normal of coherence factor for the spectral distribution
, the $\omega_k$ is random with a cut-off frequence $\omega_c=$.
The time for resuming the coherence is $t_c(k)=\frac{2\pi k}{\omega_c}$.
\ \ \\
\ \ \\
Figure 3. The probabilities $|c_1^{[L]}|$ and $|c_2^{[L]}|$ finding the
system at states $|1>$ and $|2>$ respectively.
\ \ \\
\ \ \\
Figure 4. The entropy as the function of times $L$ of measurement for
$\omega T=3\pi$, it
represent a "transition" of the information entropy from the random to
regularity. Figure 4-b is part of Figure 4-a.
\ \ \\
\ \ \\
Figure 5. The entropy as the function of times $L$ of measurement for
$\omega T=10\pi$, it
represent a "transition" of the information entropy from the random to
regularity. Figure 5-b is part of Figure 5-a.
$\omega T=10\pi$, and Figure 5-c,d with $\omega T=10\pi$.

\newpage

{\bf References} \\
\begin{enumerate}
\item J.von Neumann, Mathemstische Gruandlage de Quantumechanik, Berlin:
Springer, 1993
\item V.B.Braginski and F.Y.Khalili, {\it Quantum Measurement}, Cambridge
University Presses,1992
\item Tso-hsiu Ho, Physics (in Chinese), 22(1992) 419
\item W.H.Zurek, Phys.Today, 1991,No.10, p 36.
\item  B.Kramer,{\it Quantum Coherence in Mesoscopic Systems}, New York:
Plenum, 1991.
\item M.Gell-Mann and J.B.Hartle, Phys.Rev.D, 47(1993), 3345
\item  A.J.Leggett, S. Chakravarty, A.T. Dorsey, M.P.A. Fisher, A. Gary
and  W.Zwerger, Rev.Mod.Phys., 59(1987),1.
\item  J.P.Paz,S.Habib and W.H.Zurek, Phys.Rev.D., 47(1993),488
\item L.H.Yu and C.P.Sun, Phys.Rev.A, 49(1994),592
\item C.P.Sun and L.H.Yu, submitted to Phys.Rev.A.
\item  K.Hepp, Hev.Phys.Acta, 45(1972), 237.
\item  J.S.Bell, Hev.Phys.Acta,48(1975), 93.
\item  H.Nakazato and S.Pascazo, Phys.Rev.Lett.,70(1993),1
\item C.P.Sun, Phys.Rev.A, 48(1993),878
\item C.P.Sun, Chin.J.Phys., 32(1994),7
\item N.G.van Kampen, Physica A,153(1988), 97
\item M.Namik,S.Pascazio, Phys.Rev.A44.(1991),39.
\item M.Namik, Found. Phys.Lett,5(1992) 265
\item B.Gaveau and L.S.Schulman, J.Stat.Phys. 58(1990),1209
\item C.P.Sun,J.Phys.A,21(1988)1595.
\item C.P.Sun, Phys.Rev.D.38(1988),2908.
\item C.P.Sun, High Energy Phys.Nucl.Phys.,12(1988) 252
\item C.P.Sun,Phys.Rev.D,41(1990),1318.
\item C.P.Sun, Physica Scripta,42(1993)393
\item. B.Misra and E. Sudarshan,J.Math.Phys. 18(1977)756.
\item. W.M.Itano, O.J.Heinzen, J.J.Bollinger and D.J.Wineland,
Phys.Rev.A 41(1990)2295.
\item. T.Petrovsky,S.Tasaki and I.Prigogine. Phys.Lett. A 151(1990)109.
\item. T.Petrovsky, S.Tasaki and I.Prigogine, Phyica A 170(1991) 306.
\item. S.Inagaki, M.Namiki and T.Tajin, Phys. Lett. A 166(1992)5-12.
\item. S.Pascazio,  M.Namiki, G.Badarek, H.Ranch, Phys. Lett. A 179(1993)155.
\item. D.Home,M.Whitake, Phys. Lett. A 173(1993)327.
\item. R.Onofrio,C.Presilla and U.Tambini, Phys. Lett. A 183(1993)135.
\item. T.P.Altenmiiller and A. Schenzle, Phys.Rev. A 48(1993)70.
\item. M.J.Gagen, H.Wisemen, G.J.Milbarn, Phys.Rev. A 48(1993)132.
\item. T.P.Altenmiiller and A.Schenzle, Phys.Rev.A 49(1994)2016.
\item. G.S.Agarwal and S.P.Tewari, Phys.Lett. A 185(1994)139.
\item. A.Peres, Am.J.Phys.48(1980)931.
\item. J.Wei and E.Norman, J.Math.Phys., 4A(1963) 5
\item. M.Cini, Nuovo Cimento,73B(1983),27.
\item. T.Zaugg, M,Wikens, P.Meystre and G.Lenz
            Optics Commun. 97(1993)189.

\end{enumerate}

\end{document}